# Enhancing Dictionary Based Preprocessing For Better Text Compression

R. R. Baruah[1], V. Deka[1], M. P. Bhuyan[2]

[1](Department of Information Technology, Gauhati University, India)
[2](Department of Computer Science and Engineering, Assam Engineering College, India)

***ABSTRACT:*** With the rapid growing of data and number of applications, there is a crucial need of dictionary based reversible transformation techniques to increase the efficiency of the compression algorithms and hence contribute towards the enhancement in compression ratio. Performance analysis of compression methods in combination with the various transformation techniques is obtained for different text files of varying sizes. The popular block sorting lossless Burrows Wheeler Compression Algorithm (BWCA) is implemented along with one proposed method. For efficient compression a dictionary based transformation algorithm is also developed. It is observed that much increase in terms of compression ratio is attained when a source file is preprocessed with dictionary and then applied to BWCA and the proposed method.

## 1. INTRODUCTION

It is seen that there has been an unparalleled expansion of textual information through the use of Internet, representing text, images, video, sound, computer programs, etc. This has led to the increasing demand of the need of data compression. In computer science and information theory, data compression involves encoding information using fewer bits than the original representation. It is the ability of reducing the amount of storage or Internet bandwidth required to handle this data. The most vital objective of any compression algorithm is the compression efficiency. The more redundancy the source data has, the more effective a compression algorithm may be.

Compression can be either lossy or lossless. A lossless technique means that the restored data file is identical to the original. This is absolutely necessary for many types of data. For example: word processing files, tabulated numbers etc. Here we cannot afford to misplace even a single bit of this type of information. In comparison, data files that represent images and other acquired signals do not have to be kept in perfect condition for storage or transmission. If the changes made to these signals resemble a small amount of additional noise, no harm is done. Compression techniques that allows this type of degradation are called lossy. The distinction is important because lossy techniques are much more effective at compression than lossless methods.

The text compression techniques have captured the attention more in the recent past as there has been a substantial expansion in the usage of internet, digital storage information system, transmission of text files, and embedded system usage. Research is going on for text compression continuously to improve its methods and compressing technologies. Accordingly researchers from various parts of the world have developed compression algorithms, such as Huffman encoding, arithmetic encoding, the Lempel-Ziv family, Run length encoding based algorithms.

An alternative approach can be to develop reversible transformations that can be applied to a source text before applying any existing compression algorithms. This technique will help the back end's algorithms to compress the original text more easily which will lead to the improvement in compression ratio. These methods are advantageous in the sense that they make redundancy more visible to the compressor and are performed prior to actual compression.

The vital feature of merit for data compression is the "compression ratio", which is the ratio of the original size – compressed size divided by original uncompressed file. Thus there are two main approach to attain better compression ratio:
a) To develop different compression algorithm.
b) To develop reversible transformation that can be applied to a source text which improves the existing algorithms ability to compress. Thus preprocessing techniques came into being.

It is an important thing to note that the transformation must be exactly reversible so that the overall lossless compression paradigm is not affected. Furthermore the existing data compression and decompression algorithms are unaltered, so that they do not exploit information about the transformation while compressing. Our main objective is to improve the overall compression ratio of the original source text in comparison with what could have been achieved by using only the existing compression algorithm.

## 2. Burrows Wheeler Compression Algorithm

Within the last decade, the Burrows-Wheeler Compression Algorithm has become one of the key players in the field of universal data compression. The reasons for its success are high compression and decompression speed combined with good compression rates. BWCA is a block sorting lossless data compression algorithm which takes a block as an input. BWCA comprises of 4 stages which include Burrows Wheeler Transform (BWT),Global Structure Transformation (GST), Run Length Encoding (RLE) and Entropy Coder.

2.1 Burrows Wheeler Transform (BWT):

The fundamental concept behind this technique is that when a text file or a character string is transformed, the size of the string does not change[9]. The transformation only permutes the string into n permutations, when n is the total number of characters in the string. After performing Burrows Wheeler Transform, new transformed string can be compressed easily with compression method like run length encoding.

2.2 Global Structure Transformation(GST):

This is the second stage of the BWCA, for which MTF (Move To Front) is used. It is the most common post BWT processing algorithm which transforms the input symbol sequence into an index sequence. For each input symbol, an output index is written. The main idea is that each symbol in the data is replaced by its index in the stack of "recently used symbols". For example, long sequences of identical symbols are replaced by as many zeroes, whereas when a symbol that has not been used in a long time appears, it is replaced with a large number. Thus at the end the data is transformed into a sequence of integers; if the data exhibits a lot of local correlations, then these integers tend to be small.

2.3 Run length Encoding(RLE):

Run length encoding (RLE) is a compression algorithm which is applied when a given file contains too many redundant data or long run of similar characters. Run-length encoding performs lossless data compression and is well suited to palette-based bitmapped images such as computer icons. It does not work well at all on continuous-tone images such as photographs, although JPEG uses it quite effectively on the coefficients that remain after transforming and quantizing image blocks. Run-length encoding is used in fax machines (combined with other techniques into Modified Huffman coding). It is relatively efficient because most faxed documents are mostly white space, with occasional interruptions of black.

2.4 Huffman Coding:

Huffman coding is a data compression technique in which each input character is replaced with variable length binary digits which are called codeword and the codeword has been derived in a particular way based on the probability of occurrence of each symbol or character. The most frequent symbols in the source have the shortest length code and the least frequent symbol has the longest code. This technique is implemented by creating a binary tree of nodes. This can be stored in data structures like array or link list, the size of which depends on the number of symbols, n.

## 3. Dictionary Based Transformations:

The famous dictionary based preprocessing methods are Star Transform, Length Index Preserving Transformation (LIPT), StarNT, Intelligent Dictionary Based Encoding (IDBE) and Word Replacement Transformation (WRT). In all of them dictionary is prepared in advance and is shared by both encoder and decoder.

3.1 Star Transform

The star encoding is a reversible lossless preprocessing technique introduced by Kruse and Mukherjee [5]. The aim is to transform the text into some intermediate form which can be compressed easily by the existing data compression algorithms. The star encoding (or *-encoding) is intended to exploit the natural redundancy of the language [3]. Every word in the dictionary has a star encoded equivalent in which as many letters as possible are replaced by the "*" character.The main aim for the Star Transform is to define unique signature for each word replacing the letters of the word by a special character (*) and to use a minimum number of characters in order to identify precisely the specified word. If the word in the input text is not in the dictionary it will be passed to the backend algorithm unaltered. Hence if this transformation technique can be carried out in a proper manner, then in the resulting text we can have a more number of *characters which implies that it can be more easily compressible by the existing backend compression algorithm. Special provisions are also made in this transformation technique for handling capitalization, punctuation marks and special characters.

### 3.2 Length Index Preserving Transform

LIPT,proposed by Fauzia S. Awan and Amar Mukherjee uses not only the letters of the alphabet to denote length of the words but also to denote the offset within a block of words in the English dictionary having the same length. On creating this transformation technique, the length and the frequency of words are of utmost importance [11]. Some modifications were made to the Star transform for increasing the speed performance of the LIPT technique. In Star encoding, searching of a certain word in the encoding phase and well as in the decoding phase leads to increase in the execution time. A better solution to this problem can be attained by initially sorting in lexicographic order of the words from the dictionary and then in the encoding and decoding phase we will apply binary search in the sorted dictionary. While encoding, the symbol * denotes the beginning of the codeword, followed by the alphabets (a-z,A-Z) representing the length of the word and then the maximum of three letter codewords are places.

### 3.3 StarNT Transform

Star New Transform (StarNT), a fast transform algorithm was proposed by Weifeng Sun, Nan Zhang and Amar Mukherjee. This method is superior to LIPT not only in compression performance, but also in time complexity [10]. To gain a much better compression performance for the backend data compression algorithm, only letters [a..z, A..Z] are used to represent the codeword [12]. The first 26 words are assigned "a", "b", …,"z" as their codewords. The next 26 words are assigned "A", "B", ….. "Z". The 53rd word is assigned "aa", 54th "ab". Following this order, "ZZ" is assigned to the 2756th word in the Dictionary. The 2757th word is assigned "aaa", the following 2758th word is assigned "aab", and so on. In this transformation, the character "*" means that the following word does not exist in the transform dictionary *D*. The key reason for this change from the earlier Star family is to reduce the size of the transformed intermediate file and thus the encoding/decoding time of the backend compression algorithm can be minimized. The initial letter capitalized words and all-letter capitalized words are handled by some specialized operations.

### 3.4 IDBE Transformation:

The intelligent dictionary based encoding method was drawn by V.K. Govindan and B.S. Shajee Mohan [12]. In this transformation technique, the dictionary is produced with multiple sources of files as input. Here the codewords are formed using the ASCII characters 33 to 250. For first 218 words, the ASCII characters 33 to 250 as the code. The remaining words take each one permutation of two of the ASCII characters (in the range 33-250), in order. If there are many words left over, it can take every one permutation of three of the ASCII characters and finally if required permutation of four characters and so on. This method not only provide high compression ratio but also better security from attacks while transmission.

### 3.5 Word Replacement Transformation:

Word Replacement Transformation [13] is the most recent algorithm from the presented family. Grabowski uses only ASCII characters [128 to 255] to represent the codeword and also a promising technique invented by Taylor, trying to reduce the effect caused by end-of-line (EOL) symbols, which hamper the context, since words are usually separated by spaces. WRT gives the overall high compression ratio compared with all other preprocessing techniques using PAQ6 as the backend algorithm. WRT works based on the hashing techniques which speeds up the encoding and decoding process. The only drawback of hashing is that of the need of high memory.

3.6 Proposed algorithm while making Dictionary:

A dictionary of 5000 most frequently used words is created. The words are then arranged in decreasing order of their length i.e. highest length words are listed in the beginning of the dictionary followed by the lower length words. Then codewords are assigned to them based on the dictionary based transformation technique "StarNT." The first word is assigned as "a", second word as "b",26th word as "z",27th word as aa, 28th word as "ab",53rd word as "aaa",54th word as "aba",55th word as "aca" and so on. It is found that this generation of codewords is efficient for words that are greater than 3 because the codewords assigned to them are comparatively less than the length of the word. However problem arises when the words are of length smaller than 4 as because in that case length of the codeword is greater than the length of the word. To overcome this drawback, codewords are not assigned for those words. On encoding, these words will remain unaltered. The advantage of this technique is that for the words that are of length greater than 3, codewords of less length are obtained compared to original word means some amount of precompression is obtained before applying to any existing transformation algorithms. The words which are not found in the dictionary are stored in a temporary file and also added to the words.txt file, during encoding these words remain as they are. On making the dictionary again, the new words will be added to the dictionary and the dictionary will assign some codewords for these words.

3.7 Algorithms used:

BWCA = BWT + MTF + RLE + Huffman

Proposed Method = BWT + RLE +MTF+RLE Huffman

Dictionary Method1 = Dictionary + BWCA

Dictionary Method2 = Dictionary + Proposed Method

After observing the method based on original BWCA, we have designed one proposed method. Block diagram of all these methods are shown below. To improve the compression ratio one dictionary based transformation algorithm is designed on the basis of starNT transformation. This dictionary is also introduced in BWCA and Proposed Method.

**Method based on original BWCA:**

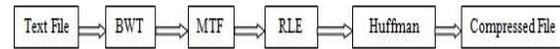

**Proposed Method:**

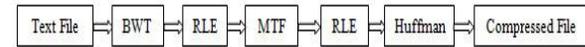

**Dictionary Method1:**

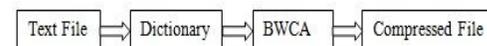

**Dictionary Method2:**

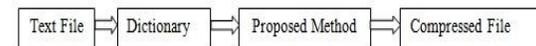

**4.EXPERIMENTAL RESULTS**
4.1 Experimental Setup

1)*Text File*: This block reads from a text file.
2)*BWT*: This block performs Burrows Wheeler Transform in the text file by taking block size of 100 bytes and writes the output in BWT.txt file.
3)*RLE*: This block performs Run Length Encoding once in the output of MTF and next in the output of BWT and writes the output in RLE.txt file.
4)*MTF* : This block performs Move To Front once in the output of BWT and next in the output of RLE and writes the output in MTF.txt file.
5)*Huffman*: This block applies Huffman coding in the output of RLE and writes the output in HUFFMAN.txt file.
6)*Dictionary*: This block performs encoding of words from the dictionary in the text file and writes the output in DICTIONARY.txt file.
7)*BWCA*: This block applies BWCA on the encoded dictionary code and writes the output in the compressed file.
8)*Proposed Method*: This block applies Proposed Method on the encoded dictionary code and writes the output in the compressed file.
9)*Compressed file*: This block contains the compressed file.

The performance analysis of the BWCA transformation algorithms along with the proposed method are done for different text files of different size. A comparison table is drawn in which comparison is done in terms of compression ratio. Implementation is done in JAVA for the different compression algorithms. The result is tabulated below.

Compression Ratio (CR) = [(Original Size − Compressed Size) ⁄ Original Size] ×100

| File (*.txt) | Original Size (bytes) | Compressed Size | | Compression Ratio | |
|---|---|---|---|---|---|
| | | Dictionary + BWCA | Dictionary + proposed method | Dictionary + BWCA | Dictionary + Proposed method |
| File1 | 289 | 144 | 200 | 50.17 | 30.8 |
| File2 | 626 | 249 | 256 | 60.22 | 59.11 |
| File3 | 1193 | 548 | 495 | 54.07 | 58.51 |
| File4 | 3341 | 1638 | 1678 | 50.97 | 49.78 |
| File5 | 4609 | 2613 | 2625 | 43.31 | 43.05 |
| File6 | 12420 | 5353 | 5408 | 56.9 | 56.46 |
| File7 | 18000 | 5916 | 6128 | 67.13 | 65.96 |
| File8 | 25808 | 11180 | 15535 | 56.68 | 39.81 |
| File9 | 34908 | 18771 | 16254 | 46.23 | 53.44 |
| File10 | 53329 | 25391 | 25952 | 52.39 | 51.34 |
| Avg. | 15452.3 | 7180.3 | 7453.1 | 53.807 | 50.826 |

Table 4.1 Size of different files before and after compression and their compression ratio with dictionary.

4.2 Different Graphs

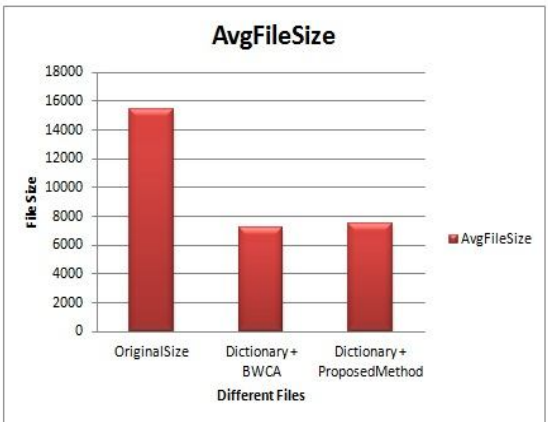

Figure 4.2.1 Average File Size

Figure shows the average file size of the original files together with dictionary with BWCA and dictionary with proposed method. Quite good amount of compression is taking place compared to original file size. Almost similar compression is seen in both the cases.

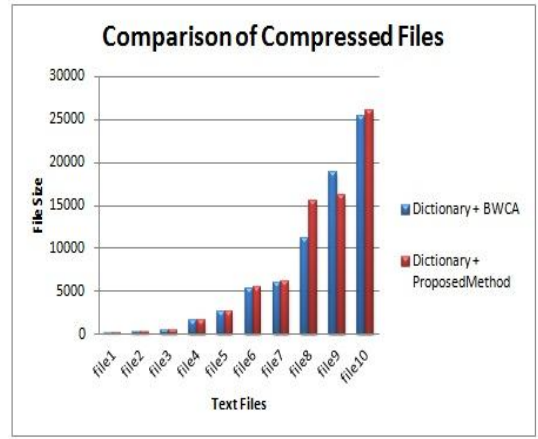

Figure 4.2.2 Comparision of Compressed Files

Fig shows the comparison of the different compressed files for Dictionary with BWCA and Dictionary with ProposedMethod. Different files are compressed at different rate for both the methods. For some files compression is showing better in case of Dictionary with BWCA whereas for other files compression is showing better in case of Dictionary with Proposed Method.

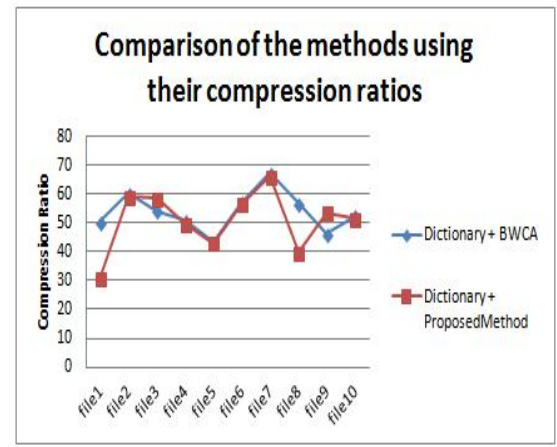

Figure 4.2.3 Comparison of the methods using their compression ratios

Figure shows the comparison of the methods dictionary with BWCA and dictionary with ProposedMethod using their compression ratios. Higher the compression ratio, more efficient is the algorithm. It is seen that different files are compressed at different rate for both the methods.

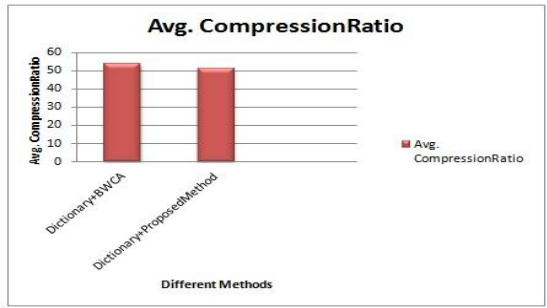

Figure 4.2.4 Average Compression ratio

Fig shows the average compression ratio of dictionary with BWCA and dictionary with proposed method. However dictionary with ProposedMethod is not showing better average compression ratio compared to dictionary with BWCA which indicates the significance of MTF immediately after BWT before applying existing compression algorithm.

## 5. CONCLUSION

After doing a detail study and various experiments on text files it is seen that application of dictionary based transformation techniques prior to transformation algorithms increases the compression ratio and thus contribute towards better compression. So the use of such type of reversible transformation technique is very useful in the field of lossless data compression. In this paper the data structure applied to the dictionary is not optimal, as in practical implementation time complexity will be more as every time dictionary needs to be rebuilt again and again. Also dictionary needs to be updated for special characters, punctuation marks etc.